\documentclass[11pt]{article}
\usepackage{hyperref}
\pdfoutput=1
\begin{document}
\title{Nuclear flow in a filamentous fungus}
\author{Patrick C. Hickey$^{1,2}$, Anna Simonin$^4$, Nick Read$^3$ \\
N. Louise Glass$^4$ and Marcus Roper$^1$\\
\\\vspace{6pt} $^1$ Dept. of Mathematics, UCLA, Los Angeles, CA 90095, USA \\
$^2$ NIPHT Ltd., Edinburgh, EH9 3JF, UK \\
$^3$ Institute of Cell Biology, University of Edinburgh, \\ Edinburgh EH9 3JH, UK \\
$^4$ Dept. of Plant and Microbial Biology, UC Berkeley, \\ Berkeley, CA 94720, USA
}

\maketitle
\begin{abstract}
The syncytial cells of a filamentous fungus consist of a mass of growing, tube-like hyphae. Each extending tip is fed by a continuous flow of nuclei from the colony interior, pushed by a gradient in turgor pressure. The myco-fluidic flows of nuclei are complex and multidirectional, like traffic in a city. We map out the flows in a strain of the model filamentous fungus {\it N. crassa} that has been transformed so that nuclei express either hH1-dsRed (a red fluorescent nuclear protein) or hH1-GFP (a green-fluorescent protein) and report our results in a fluid dynamics video.
\end{abstract}
% main text
\section{Introduction}

The filamentous fungi are the most diverse of any eukaryotic organisms, thriving as mutualists, decomposers and pathogens the world over. One suspected contributor to their tremendous ecological success is their unusual mode of life. Unlike plant and animal cells, the cells of filamentous fungi are generally multinucleate, and can even harbor genetically different nuclei, bathed by a common cytoplasm. As the tube-like hyphae grow, each extending incrementally at its tips, nuclei flow from the colony interior to fill the free space created at the tips. We show that the flow is driven by pressure gradients across the colony, and that nuclei follow complex multi-directional trajectories, reminiscent of cars traveling through a city. We hypothesize that the complexity of nuclear paths is a deliberate effort by the fungus to keep genetically different nuclei well mixed. Disrupting the exquisite hydraulic engineering of the cell (e.g. by knocking out the ability of hyphae to fuse, to make the multi-connected network) causes genetically different nuclei to become un-mixed during growth.

\section{Discussion}

Our fluid dynamics video includes 11 short segments:

\begin{enumerate}
\item A time lapse sequence, accelerated 7500 fold showing a colony invading a small block of agar.
\item Confocal imaging of hH1-GFP transformed nuclei flowing toward a growing tip. As we pan deeper into the colony, we see how these tip hyphae are fed as branches of trunk hyphae, each supplying nuclei to many tips. The flow speed in a tip hyphae is $\sim$ 0.1 $\mu$m/s, flow speed in trunk hyphae can be 30 or 80 times greater.
\item In the colony interior, nuclei flowing through the complex network of hyphae follow torturous and even multidirectional paths, looking in confocal microscopy, like the headlights of cars navigating a microscopic city. Flow speeds in the colony interior can reach 10-20 $\mu$m/s.
\item Just like cars in a city, speeds vary between hyphal roads. Some roads remain grid-locked while in others nuclei flow at speeds of up to 10-20 $\mu$m/s.
\item Under some conditions, in small colonies, the nuclei form spontaneous traffic jams.
\item What drives the fluid flow? We measured the variation in nuclear speeds across hyphae. When collapsed by hyphal diameter and axial flow speed, we see a common Poiseuille-flow profile in each hypha, indicating that the flows are hydrodynamic: driven by pressure gradients across the colony.
\item We can manipulate the pressure gradients by applying hyper-osmotic solutions to the colony. These treatments apply a uniform pressure gradient counter to the normal direction of flow. Flow in each fungal hyphae is transiently reversed, indicating that the hyphal network architecture -- far from being random -- is finely tuned to create mixing flows from spatially coarse pressure gradients.
\item What is the function of these flows? Since real fungi can harbor many genetically different nucleotypes, we hypothesize that physical mixing, associated with the flows, helps to preserve nuclear diversity as the fungus grows. We show a sequence of nuclear mixing in a chimeric colony formed by fusing two {\it N. crassa} colonies; one with hH1-dsRed and one with hH1-GFP transformed nuclei. 
\item We can manipulate the hyphal architecture genetically. We show micrographs of conidiophores (spore-bearing hyphae) of chimeric wild type and {\it so} mutant colonies. Both colonies have dsRed (red) and GFP (green) expressing nucleotypes. In the wild type colony, the red and green nuclei remain well mixed in the colony. In the {\it so} colony they segregate out, so that conidiophores end up almost exclusively red or exclusively green (shown). {\it so} mutations stop the hyphae from fusing to create an interconnected network.
\item Conidiophores are branched structures (without interconnections) so pose particular challenges for mixing. Taylor dispersion (enhanced diffusion due to fluid shear across a hypha) can keep nuclei well-mixed but only if flow speeds in the conidiophore exceed 200 $\mu$m/s. To achieve such fast flows each conidiophore must feed into at least 2000 hyphal tips.
\item A time-lapse video showing the growth and collapse of conidiophores, accelerated 7500 fold. 2000 separately growing hyphae places impossible weight upon the conidiophore, causing it to collapse. Since spores are thought to be dispersed by air-flows across the colony and collapsed conidiophores are likely to experience reduced wind speeds, rates of spore liberation are reduced. Loss of spore dispersal effectiveness represents the high price paid by the colony for maintaining mixing flows.
\end{enumerate}
\end{document}